\documentclass[prd,twocolumn,showpacs,preprintnumbers,aps,nofootinbib]{revtex4-1}

\usepackage[dvipdfmx]{graphicx}  
\usepackage{dcolumn}
\usepackage{bm}
\usepackage{amsmath,amssymb,amsfonts}
\usepackage{latexsym}
\usepackage{color}

\def\nn    {\nonumber}

\begin{document}

\title{\boldmath Top-Assisted Di-Higgs boson Production Motivated by Baryogenesis}
\author{Wei-Shu Hou, Masaya Kohda and Tanmoy Modak}
\affiliation{Department of Physics, National Taiwan University, Taipei 10617, Taiwan}
\bigskip


\begin{abstract}
We study top-assisted di-Higgs production via $cg \to tH \to thh$, 
where $h$ is the 125 GeV scalar boson, and $H$ is the $CP$-even heavy Higgs.
The context is the two Higgs doublet model without a $Z_2$ symmetry, 
where the extra Yukawa coupling $\rho_{tc}$ generates $tH$ production,
with the extra top Yukawa $\rho_{tt} \simeq 0$ to avoid $gg \to H$ constraints.
We find that discovery is possible for $m_H$ around 300 GeV or so at the LHC,
but would need finite $h$-$H$ mixing angle $\cos\gamma$
to allow for finite $\lambda_{Hhh}$ coupling,
and $\rho_{tc}$ also needs to be not too small. 
A sizable $\rho_{tc}$ could drive electroweak baryogenesis, 
which further motivates the search.
\end{abstract}

\maketitle

\section{Introduction}

The highlight at the Large Hadron Collider (LHC) so far is the discovery of
the 125 GeV scalar boson $h$~\cite{h125_discovery} in 2012,
which resembles rather closely~\cite{Khachatryan:2016vau} 
the Higgs boson of the Standard Model (SM).
To improve our understanding of the Higgs potential,
a key goal at the LHC is to search for di-Higgs, or $pp \to hh$, production.
The program is rather challenging, as $hh$ production in 
SM~\cite{Grazzini:2018bsd,twikidihiggs} is rather suppressed. 
The ATLAS~\cite{Aaboud:2018knk,Aaboud:2018ftw,Aaboud:2018ewm} 
and CMS~\cite{Sirunyan:2017djm,Sirunyan:2017guj,Sirunyan:2018iwt,CMS:2018smw}
experiments have already conducted searches for 
resonant and non-resonant di-Higgs production,
but there is little expectation that the SM process can be observed
even at the High-Luminosity LHC (HL-LHC). 
In this paper we explore a novel possibility with New Physics,
that of resonant $hh$ production in association with a top quark.

The context is a two Higgs doublet model (2HDM) with {\it extra Yukawa couplings},
i.e. without a discrete $Z_2$ symmetry to impose the Glashow-Weinberg~\cite{Glashow:1976nt}
NFC (Natural Flavor Conservation) condition to forbid flavor-changing neutral Higgs (FCNH) couplings.
Note that the usual $Z_2$ symmetry eliminates {\it all} extra Yukawa couplings.
Two processes can be operative that feed di-Higgs production. 
The existence of extra diagonal Yukawa coupling $\rho_{tt}$ of 
the exotic $CP$-even neutral Higgs boson $H$ means that
one could have $gg \to H \to hh$ production through triangle top quark loop.
A second process depends on the FCNH $tcH$ coupling, $\rho_{tc}$,
whereby one can have $cg\to t H \to t hh$ (conjugate process implied). 
In a previous paper~\cite{Hou:2018zmg}, we considered turning $\rho_{tt}$ off, so the 
first process is subdued, and one is left with the second process:
di-Higgs and top associated production. In this paper we focus on this process, exploiting the extra top quark
to investigate possible prospects at the LHC.

We stress that 2HDM without extra $Z_2$ symmetry could~\cite{Fuyuto:2017ewj}  
account for baryon asymmetry of the Universe (BAU), 
via electroweak baryogenesis (EWBG).
The leading mechanism is via $\rho_{tt}$ and is rather robust.
However, in exploring~\cite{Hou:2018zmg} the prospect for a lighter pseudoscalar $A$ boson
around 300 GeV, in face of direct search bounds, we opted to turn off $\rho_{tt}$,
noting that $\rho_{tc}\sim 1$ would still offer 
an alternative mechanism~\cite{Fuyuto:2017ewj}  for EWBG,
hence is interesting in itself.
Thus, the $\rho_{tc}$ driven $cg\to t H\to t hh$ process studied here
is a companion to the $cg\to t A \to tt\bar c$ process
that bears a rather intriguing signature.

We find that $thh$ discovery is possible at the HL-LHC
for relatively light $H$, 
where the associated top quark gives extra handle on background reduction.
However, a relatively large $Hhh$ coupling would be needed,
hence the prospect cannot be said as very likely, but it is not negligible.
In the following, we start with the formalism in Sec.~\ref{form}, then the collider
signatures in Sec.~\ref{coll} and end with some discussions in Sec.~\ref{dissc}.

\section{Formalism}\label{form}

The $CP$-even scalars $h$, $H$ and $CP$-odd scalar $A$ couple to fermions 
by~\cite{Davidson:2005cw, Altunkaynak:2015twa}
\begin{align}
&-\frac{1}{\sqrt{2}} \sum_{F = U, D, L}
 \bar F_{iL} \bigg[\big(-\lambda^F_{ij} s_\gamma + \rho^F_{ij} c_\gamma\big) h \nn\\
 &+\big(\lambda^F_{ij} c_\gamma + \rho^F_{ij} s_\gamma\big)H -i ~{\rm sgn}(Q_F) \rho^F_{ij} A\bigg]  F_{jR}
   +{\rm h.c.},
\end{align}
where $i,j =1,2,3$ are generation indices that are summed over, 
$\lambda^F_{ij}=({\sqrt{2}m_i^F}/{v})\, \delta_{ij}$  (with $v \simeq 246$ GeV) 
and $\rho^F$ are $3\times 3$ real diagonal and complex matrices, respectively.
With shorthand $c_\gamma= \cos\gamma$, $s_\gamma= \sin\gamma$,
the mixing angle $\gamma$ is usually written as $\alpha-\beta$ in Type-II 2HDM notation. 
However, as we advocate no $Z_2$ symmetry
and there exists a second set of Yukawa couplings $\rho^F_{ij}$,
we prefer the notation of Ref.~\cite{Hou:2017hiw},
since $\tan\beta$ is ill-defined. 
The FCNH couplings of interest for $tc H$ are 
$\rho^U_{23}\equiv \rho_{ct}$ and $\rho^U_{32}\equiv \rho_{tc}$. 
$B$ physics sets stringent limits on $\rho_{ct}$~\cite{Altunkaynak:2015twa},
while $\rho_{tc}$ is only mildly constrained~\cite{Crivellin:2013wna},
depending on $m_{H^+}$. In our study, we set $\rho_{ct} = 0$ and 
take $|\rho_{tc}| < 1$.

The most general $CP$-conserving two Higgs doublet potential is
given in Higgs basis as~\cite{Davidson:2005cw, Hou:2017hiw}
\begin{align}
 & V(\Phi,\Phi') = \mu_{11}^2|\Phi|^2 + \mu_{22}^2|\Phi'|^2 - (\mu_{12}^2\Phi^\dagger\Phi' + h.c.)
 \nn\\
 & \quad + \frac{\eta_1}{2}|\Phi|^4 + \frac{\eta_2}{2}|\Phi'|^4 + \eta_3|\Phi|^2|\Phi'|^2
              + \eta_4 |\Phi^\dagger\Phi'|^2\nn\\
 & + \bigg[\frac{\eta_5}{2}(\Phi^\dagger\Phi')^2
     + \left(\eta_6 |\Phi|^2 + \eta_7|\Phi'|^2\right) \Phi^\dagger\Phi' + h.c.\bigg],
\label{pot}
\end{align}
where $v$ arises from the doublet $\Phi$ via $\mu_{11}^2=-\frac{1}{2}\eta_1 v^2$, 
while $\left\langle \Phi'\right\rangle =0$ (hence $\mu_{22}^2 > 0$), 
$\eta_i$s are quartic couplings, 
again in the notation of Ref.~\cite{Hou:2017hiw}.
A second minimization condition,
$\mu_{12}^2 = \frac{1}{2}\eta_6 v^2$, removes $\mu_{12}^2$
and  reduces the total number of parameters to nine~\cite{Hou:2017hiw}.
The mixing angle $\gamma$ between the $CP$ even bosons
satisfies the {relations}~\cite{Hou:2017hiw}
\begin{align}
 c_\gamma^2 = \frac{\eta_1 v^2 - m_h^2}{m_H^2-m_h^2},~\quad \quad \sin{2\gamma} = \frac{2\eta_6 v^2}{m_H^2-m_h^2},
\end{align}
which, for $c_\gamma$ small but not infinitesimal, 
one has $ c_\gamma \simeq|\eta_6| v^2/(m_H^2 -m_h^2)$.
This is approximate alignment~\cite{Hou:2017hiw}, 
i.e. small $c_\gamma$ values can be attained without 
requiring $\eta_6$ to be small.
But in the alignment {\it limit}, $c_\gamma \to 0$, 
either~\cite{Hou:2017hiw} $\eta_6$ has to vanish (and $m_h^2 \to \eta_1 v^2$), 
or else one has decoupling~\cite{Gunion:2002zf}, 
i.e. $m_H^2/v^2 \gg 1$.

We are interested in the $Hhh$ coupling, which is the coefficient of the
$\lambda_{Hhh}H h^2$ term derivable from Eq.~\eqref{pot}, 
\begin{align}
\lambda_{Hhh} & =  \frac{v}{2} \bigg[3 c_\gamma s_\gamma^2 \eta_1+ c_\gamma (3c_\gamma^2 - 2) \eta_{345}\nn\\
 & \quad\quad + 3 s_\gamma( 1-3c_\gamma^2)\eta_6+3s_\gamma c_\gamma^2 \eta_7\bigg]\label{lHhh},
\end{align}
with $\eta_{345}=\eta_3+\eta_4+\eta_5$.
It reduces further to
\begin{align}
 \lambda_{Hhh}\simeq\frac{c_\gamma}{2}v\bigg[3\frac{m_H^2}{v^2}-2\eta_{345}+3\mbox{sgn}(s_\gamma) 
 c_\gamma \eta_7+\mathcal{O}(c_\gamma^2)\bigg]\label{lHhhapp},
\end{align}
for small $c_\gamma$,
so $\lambda_{Hhh} \to 0$ as $c_\gamma\to 0$.
To enhance $cg\to tH\to t hh$, sizable $\lambda_{Hhh}$ is needed,
and $\eta_{345} <0$ may be preferred so the first two terms add up.
However, $\lambda_{Hhh}$ could still be sizable if 
$2 \eta_{345} \gg {3 m_H^2}/{v^2} > 0$.
Either way, a large $|\eta_7|$ with proper sign for $\eta_7c_\gamma s_\gamma$ would  help.

The quartic couplings $\eta_1$, $\eta_{3{\rm -}6}$ can be expressed in 
terms of $m_h$, $m_A$, $m_H$, $m_{H^\pm}$, $\mu_{22}$,
all normalized to $v$, as well as the mixing angle $\gamma$~\cite{Hou:2017hiw}:
\begin{align}
 & \eta_1 = \frac{m_h^2 s_\gamma^2 + m_H^2 c_\gamma^2}{v^2},\\
 & \eta_3 =  \frac{2(m_{H^\pm}^2 - \mu_{22}^2)}{v^2},\\
 & {\eta_4 = \frac{m_h^2 c_\gamma^2 + m_H^2 s_\gamma^2 -2 m_{H^\pm}^2+m_A^2}{v^2}},\\
& \eta_5 =  \frac{m_H^2 s_\gamma^2 + m_h^2 c_\gamma^2 - m_A^2}{v^2},\\
 & \eta_6 =  \frac{(m_h^2 - m_H^2)(-s_\gamma)c_\gamma}{v^2},
\end{align}
but $\eta_2$ and $\eta_7$ are not related to masses, nor the mixing angle $\gamma$. 
Thus, we take $v$, $\gamma$, $m_h$, $m_A$, $m_H$, $m_{H^\pm}$, $\mu_{22}$, $\eta_2$ and $\eta_7$ 
as the phenomenological parameters.
To save computation time, we randomly generate these parameters 
in the following ranges:
$\mu_{22} \in [0, 700]$  GeV, 
$m_H \in [250, 500]$ GeV, 
$m_{H^\pm} \in [300, 600]$ GeV,
$\eta_2 \in [0, 3]$, $ \eta_7 \in [-3, 3]$, 
and $\gamma$ values that satisfy
 $c_\gamma \in [0, 0.2]$, 
with $m_h = $  125 GeV.
{We choose two different scenarios for $m_{A}$.
In the main scenario, we generate $m_{A}\in [250, 500]$ GeV,  with $m_H < m_A,~m_{H^\pm}$. 
In the second scenario we take $|m_A- m_H|<\delta$,
where the choice of $\delta$ is discussed later in the section.}
We explore up to $m_H = 500$ GeV because,
while $\lambda_{Hhh}$ and $\mathcal{B}(H\to hh)$ increases
with $m_H$, the discovery potential for $cg\to tH\to t hh$ 
suffers the drop in parton luminosities for heavier $m_H$.

The dynamical parameters in the Higgs potential, Eq.~\eqref{pot}, 
need to satisfy perturbativity, tree-level unitarity and positivity conditions, 
for which we utilize 2HDMC~\cite{Eriksson:2009ws}.
2HDMC uses the input parameters~\cite{Eriksson:2009ws} 
$m_{H^\pm}$ and $\Lambda_{1-7}$ in Higgs basis, and with $v$ implicit. 
We identify $\Lambda_{1-7}$ with $\eta_{1-7}$.
Further, we {conservatively} demand all $|\eta_i| \leq 3$, while 
$\eta_2 >0$ is required by the potential positivity, in addition to more
involved conditions for other couplings.
To match the convention of 2HDMC, we take $-\pi/2\leq \gamma \leq \pi/2$.

\begin{figure*}[t!]
\center
\includegraphics[width=.33 \textwidth]{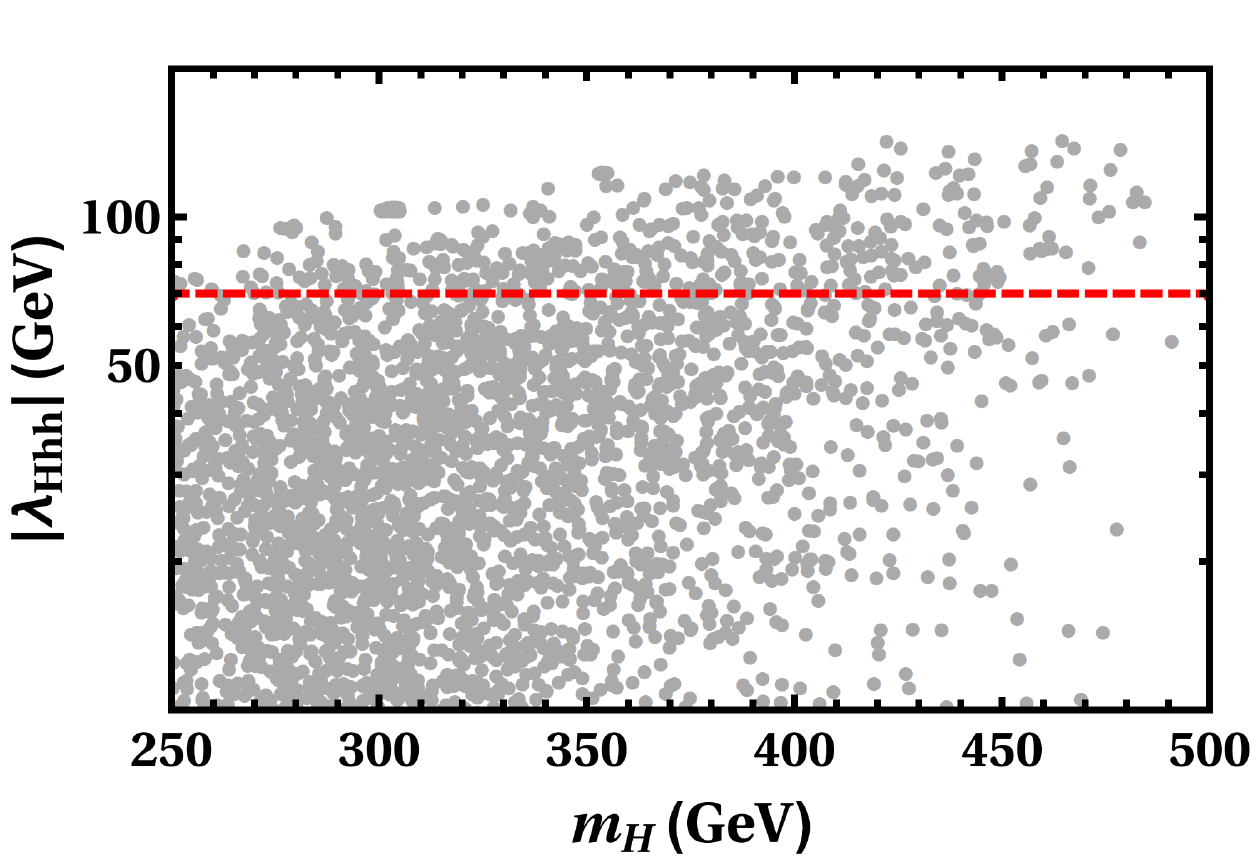}
\includegraphics[width=.33 \textwidth]{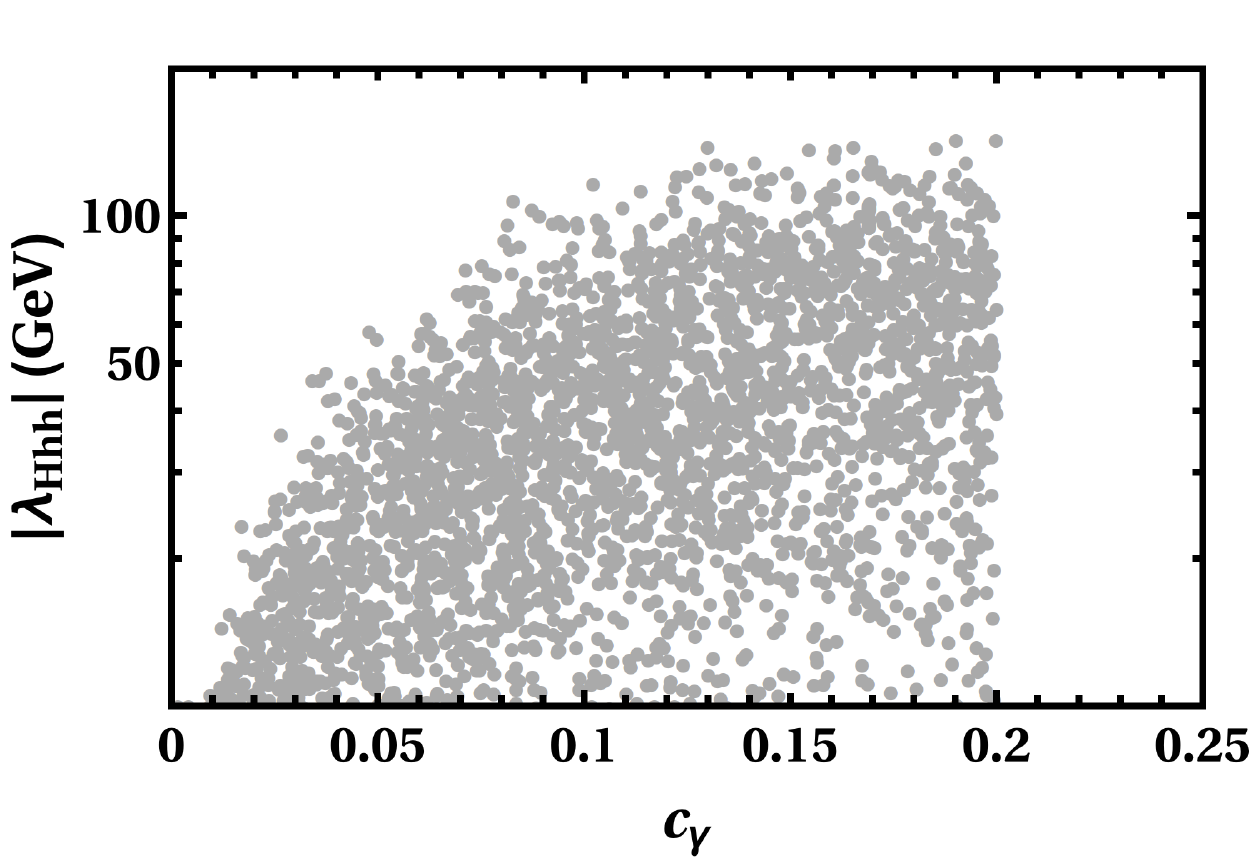}
\includegraphics[width=.33 \textwidth]{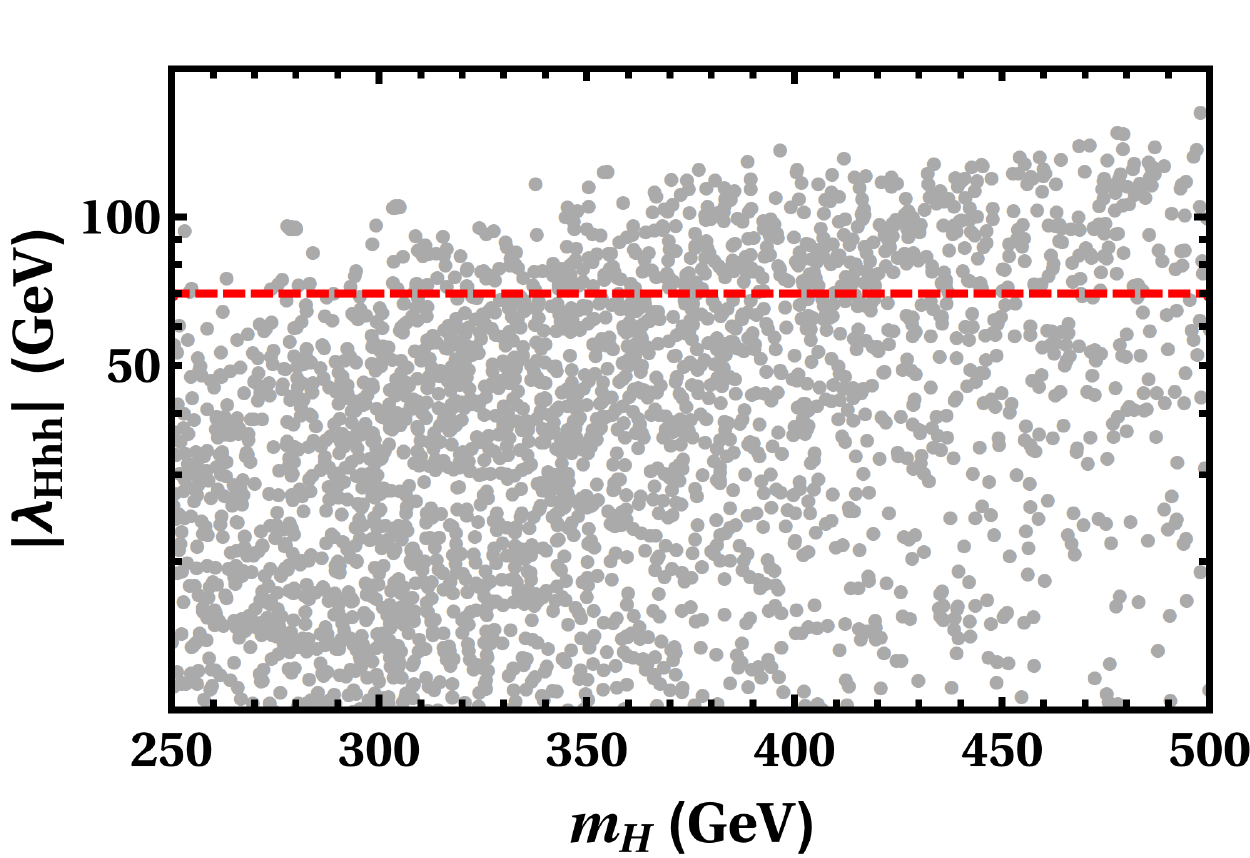}
\includegraphics[width=.33 \textwidth]{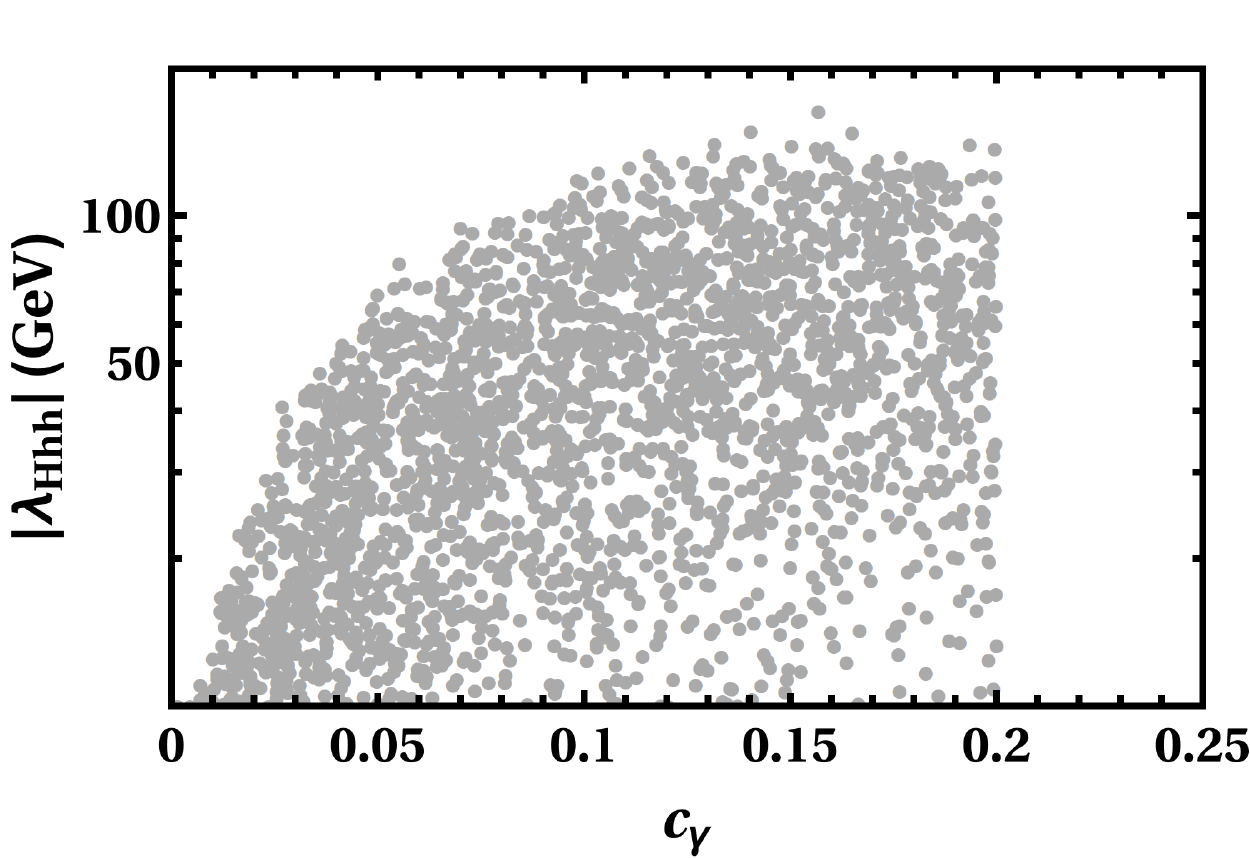}
\caption{
The $\lambda_{Hhh}$ vs $m_H$ and $c_\gamma$ plots for 
the scan points that pass perturbativity, tree-level unitarity 
and positivity through 2HDMC, where $|\eta_i| < 3$ is maintained.
The $T$ parameter constraint is also imposed.
Upper panels are for $m_H < m_A, m_{H^\pm}$, 
and lower panels are for $|m_A- m_H|< 5$ GeV. 
See text for detailed explanation.
} 
\label{lH0hh}
\end{figure*}

\begin{table*}[t]
\centering
\begin{tabular}{c |c| c| c| c | c | c| c | c |c| c|c| c| c| c|c}
\hline
BP  & $\eta_1$ &  $\eta_2$   &  $\eta_3$   & $\eta_4$  & $\eta_5$ &$\eta_{345}$ & $\eta_6$  & $\eta_7$  & $m_{H^\pm}$  & $m_A$ & $m_H$ & $c_\gamma$  
&$s_\gamma$ & $|\lambda_{Hhh}|$ & $\frac{\mu_{22}^2}{v^2}$\\
 &&&&&&&&& (GeV) & (GeV) & (GeV) &&& (GeV)&\\ 
\hline
\hline
1         & 0.287& 2.996& $-0.188$ & 2.039& $-2.555$ & $-0.704$ & $-0.172$ & 0.557& 303.27& 480.96& 279.49& 0.169& $-0.986$ & 96.80& 1.61 \\
2         & 0.294 & 2.781& 0.269& 2.095& $-2.945$ & $-0.581$ & $-0.21$ & 0.633& 340.18& 518.26& 303.48& 0.169 & $-0.986$ & 104.32& 1.77     \\

3         & 0.309& 2.984& $-0.017$ & 2.416& $-2.727$ & $-0.328$ & $-0.301$ & 0.881& 362.90& 536.38& 354.15& 0.169 & $-0.986$ & 123.03& 2.18\\
&&&&&&&&&&&&&&&\\
$a$       & 0.286& 2.97& 1.508& $-2.189$ & $-0.02$ & $-0.701$ & $-0.169$ & 0.525& 377.31& 276.72& 277.71& 0.169& $-0.986$ & 96.07& 1.59\\
$b$       & 0.294& 2.973& 1.42& $-1.976$ & $-0.037$ & $-0.593$ & $-0.211$ & 0.631& 388.90& 304.0& 303.92& 0.169 & $-0.986$ & 105.06& 1.78\\
$c$       & 0.309& 2.976& 0.294& $-0.62$ & $-0.045$ & $-0.371$ & $-0.301$ & 0.932& 377.17& 353.3& 353.89& 0.169& $-0.986$ & 124.00& 2.2\\
%
\hline
\hline
\end{tabular}
\caption{Parameter values for the six benchmark points of Table~\ref{bench}. See text for details.}
\label{bench}
\end{table*}

We also need to impose the stringent oblique $T$ parameter~\cite{Peskin:1991sw} constraint,
which constrains the scalar masses $m_H$, $m_A$ and $m_{H^\pm}$~\cite{Froggatt:1991qw,Haber:2015pua}, 
and hence $\eta_i$s. 
We apply the $T$ parameter constraint~\cite{Baak:2013ppa} on 
the points that passed 2HDMC, using the expression given in Ref.~\cite{Haber:2015pua}.
The final ``scanned points'' within $2\sigma$ error of $T$ parameter
are plotted in Fig.~\ref{lH0hh}. 
The upper panel is for $m_H < m_{H^\pm}$, $m_A$,
such that $H\to A Z,~H^\pm W^\mp$ decays are disallowed, 
which in turn enhances $H\to h h$ branching ratio.
As expected, the upper range for $\lambda_{Hhh}$ mildly increases 
as $m_H$ becomes heavier, but vanishes with $c_\gamma \to 0$.

The FCNH coupling $\rho_{tc}$ also receives constraint from CMS four 
top search~\cite{Sirunyan:2017roi} through the $cg\to t H\to t t \bar c$ process, 
{which  is proportional to 
$\left|s_\gamma \rho_{tc}\right|^2$~\cite{Hou:2018zmg} if we take $\rho_{ct}=0$}.
However, if  $A$ and $H$ are mass and width degenerate,
the processes $cg\to t A\to t t \bar c$ and $cg\to t H\to t t \bar c$ 
cancel each other exactly~\cite{Hou:2018zmg, Kohda:2017fkn},
resulting in potentially much weaker constraint on  $\rho_{tc}$, 
which can in principle give rise to larger $thh$ production.
The lower panels of Fig.~\ref{lH0hh} are for this scenario of 
nearly degenerate $A$ and $H$, where we assume 
$|m_A- m_H|< \delta$, with $\delta = 5$ GeV for illustration. 
The dependence of $\lambda_{Hhh}$ on $m_H$ and $c_\gamma$ 
is similar as in $m_H< m_{H^\pm}, m_A$ case.  
We note that in the left panels,
i.e. $\lambda_{Hhh}$ vs $m_H$, we have
drawn a line at $\lambda_{Hhh} \simeq 70$ GeV,
to illustrate that $\lambda_{Hhh}$ can be sizable
over a finite parameter region.

\section{Collider signature}\label{coll}
The discovery potential of the $cg\to tH \to t h h$ (conjugate process implied) 
depends on the $hh$ decay final states. 
In this paper, we primarily focus on $t\to b \ell^+ \nu_{\ell}$ ($\ell = e,\mu$) with
both $h$ bosons decaying via $h \to b\bar b$, giving rise to five $b$-jets, 
one lepton and missing transverse energy ($E_T^{\rm miss}$) signature. 
We do not look for hadronic decay of 
$t$ due to QCD multi-jet backgrounds, as also discussed in~\cite{Buschmann:2016uzg}.
In general, $h h\to \gamma \gamma b \bar b$ and $h h\to \tau\bar  \tau b \bar b$ 
modes are suppressed.
But $h h \to WW^* b \bar b$ decay could provide some sensitivity, 
which we discuss towards the end of this section.

We set all $\rho_{ij}=0$ except $\rho_{tc}$ for simplicity throughout this section.
Due to the presence of non-zero $c_\gamma$, 
the branching ratios of $h$ will be modified compared with SM, 
albeit in minor way. 
Since we set all $\rho_{ij}=0$ except $\rho_{tc}$ 
and assume $|c_\gamma| <0.2$, 
the branching ratios remain practically the same. 
In the following, we assume 
all branching ratios of $h$ are SM-like for simplicity.

To illustrate the discovery potential of $cg\to tH \to t h h$, 
we choose six benchmark points (BP) from Fig.~\ref{lH0hh} 
with large $|\lambda_{Hhh}|$ values, which are summarized in Table~\ref{bench}.  
The first three, BP1, BP2 and BP3, are for 
the $m_H< m_{H^\pm}, m_A$ scenario, 
while the other three, BP$a$, BP$b$ and BP$c$, 
are for $|m_A- m_H|< 5$ GeV scenario. 
The values of $m_H$ are chosen for 
$2 m_h < m_H < 300$ GeV,  $m_H\approx 300$ GeV, 
and $m_H> 2 m_t$ (above the $t\bar t$ threshold), respectively.
All six benchmark points are for $\eta_{345} < 0$, in accordance with 
the discussions in the preceding section to achieve large $|\lambda_{Hhh}|$.

\begin{table}[b]
\centering
\begin{tabular}{c| c| c| c| c | c }
\hline
BP  & \hspace{.3cm} $\rho_{tc}$    \hspace{.3cm} &  \hspace{.3cm}$tc$  \hspace{.3cm}& \hspace{.3cm}  $h h$  \hspace{.3cm}&  \hspace{.3cm} $WW$  
\hspace{.3cm}&  \hspace{.3cm}$ZZ$  \hspace{.3cm} \\
\hline
\hline
     1    & 0.54      & 0.698  & 0.232    & 0.049   & 0.021 \\
     2    & 0.54      & 0.688  & 0.238    & 0.051   & 0.023 \\
     3   & 0.54       & 0.677   & 0.235  & 0.06    & 0.027\\
&&&&&\\  
   $a$    & 0.54      & 0.700  & 0.229    & 0.049   & 0.021 \\
   $b$    & 0.54      & 0.686  & 0.240    & 0.051   & 0.023 \\
   $c$    & 0.54      & 0.674  & 0.238    & 0.059   & 0.027 \\     
\hline
\hline
\end{tabular}
\caption{$H$ decay branching ratios for the benchmark points.}
\label{branch}
\end{table}

The $cg\to tH \to t h h$ process depends also on $\rho_{tc}$.
For sizable $c_\gamma$, the available parameter space 
for $\rho_{tc}$ is constrained by the $\mathcal{B}(t\to c h)$ measurement. 
The latest ATLAS 95\% CL upper limit (with 13 TeV 36.1 fb$^{-1}$ data)
is $\mathcal{B}(t\to c h) < 1.1\times 10^{-3}$~\cite{Aaboud:2018oqm}.
Using this limit and our $c_\gamma$ value, 
we find the upper limit on $\rho_{tc} =0.54$, applicable to all six benchmark points. 
We find the $2\sigma$ upper limit
on $\rho_{tc}$ for BP1, BP2 and BP3 to be 0.56, 0.55, 0.63 respectively. 
The BP$a$, BP$b$ and BP$c$ benchmark points were chosen such that
the constraint from Ref.~\cite{Sirunyan:2017roi} becomes much weaker
due to cancellation between $cg\to t A\to t t \bar c$ and $cg\to t H\to t t \bar c$~\cite{Hou:2018zmg, Kohda:2017fkn}. 
However, besides the aforementioned $t \to ch$ constraint,
$\rho_{tc}$ can still be constrained by $B_{s,d}$ mixing and $\mathcal{B}(B\to X_s\gamma)$, where $\rho_{tc}$ enters
via charm loop through $H^+$ coupling~\cite{Crivellin:2013wna, Altunkaynak:2015twa}.
A reinterpretation of the result from Ref.~\cite{Crivellin:2013wna},
finds $|\rho_{tc}|\lesssim 1.7$ for $m_{H^\pm}=500$ GeV~\cite{ Altunkaynak:2015twa}. 
In our analysis we choose $\rho_{tc}=0.54$ for all six benchmark points,
where the $H$ decay branching ratios are given in Table~\ref{branch}~\footnote{Note that for BP3 and BP$c$ the coupling 
$\lambda_t c_\gamma$ induces $ H \to t \bar t $ decay. However,
for both of the benchmark points $\mathcal{B}(H \to t \bar t)\lesssim 0.001$ and not displayed in Table~\ref{branch}.}.

We remark that the FCNH $tuH$ coupling\footnote{
$\rho_{ut}$ is tightly constrained by $B_d$ mixing and $b \to d \gamma$~\cite{Crivellin:2013wna}.}
$\rho_{tu}$ can also induce top-assisted di-Higgs
via $ug\to tH \to t h h$, and our analysis can be extended to the case 
where all $\rho_{ij} = 0$ except $\rho_{tu}$ (see also Ref.~\cite{Buschmann:2016uzg}).
While the ATLAS 95\% CL upper limit $\mathcal{B}(t\to u h)< 1.2\times 10^{-3}$~\cite{Aaboud:2018oqm}
is not  much different from the $t \to ch$ case,
the CMS four top search~\cite{Sirunyan:2017roi} would give a stronger limit 
on $\rho_{tu}$ than the $\rho_{tc}$ case. 
The latter is because the relevant process $qg\to t H\to t t \bar q$ ($q = u, c$) is enhanced by 
the parton distribution function (PDF) of up quark while the signal region does not 
differentiate $\bar u$ and $\bar c$.
Similarly and more efficiently, the $t$-channel scalar exchange process $qq \to tt$ via $\rho_{tq}$ is 
enhanced by up PDF; 
hence, the ATLAS same-sign top search~\cite{Aad:2015gdg} may provide a significant constraint 
in contrast to the $\rho_{tc}$ case~\cite{Hou:2018zmg}.
Despite stronger constraints on $\rho_{tu}$, the discovery potential of $ug \to tH \to t hh$ 
would be balanced to some extent by the similar up-PDF enhancement in comparison with the $\rho_{tc}$ case;
but, we do not expect improvement in the signal significance for BP1, BP2 and BP3.
By contrast, for an equivalent of BPa, BPb or BPc, $ug\to t A\to t t \bar u$ effectively cancels 
$ug\to t H\to t t \bar u$, relaxing the four top search limit.
The same is true for the ATLAS $qq \to tt$ limit.
In such a case, under the $t \to u h$ constraint, $\rho_{tu}$ may be as large as the $\rho_{tc}$ case;
hence, we expect a better discovery potential of $ug \to tH \to t hh$, boosted by up PDF.

To investigate the discovery potential of top assisted di-Higgs production 
at the LHC, we study $pp\to t H +X \to t h h +X$ with 
both $h$ decaying to $b\bar b$, while $t\to b \ell^+ \nu_{\ell}$. 
The dominant backgrounds are
$t\bar t+$jets, 
single-top, 
$t\bar t h$, $4t$, 
$t\bar t W$ and $t\bar tZ$, while 
$tZj$, DY+jets, $W+$jets and $tWh$ are subdominant. 
We  do not include backgrounds from non-prompt and fake sources, 
as these are not properly modeled in Monte Carlo simulations 
and require data to estimate.
We generate signal and background event samples at LO, utilizing 
Monte Carlo event generator MadGraph5\_aMC@NLO~\cite{Alwall:2014hca}
with default PDF set NN23LO1~\cite{Ball:2013hta}
for $pp$ collisions at $\sqrt{s}=14$ TeV, interfaced with 
PYTHIA~6.4~\cite{Sjostrand:2006za} for showering and hadronization,
and adopt MLM matching scheme~\cite{Alwall:2007fs}
for matrix element and parton shower merging.
The event samples are then fed into Delphes~3.4.0~\cite{deFavereau:2013fsa}
for detector effects (ATLAS based). The effective Lagrangian is implemented using FeynRules~2.0~\cite{Alloul:2013bka}.

\begin{table}[t!]
\centering
\begin{tabular}{c |c| c| c c }
\hline
                     BP     &  \ Signal \       &  \ Total Bkg.  \      &  \ Significance      \\ 
                            &       (fb)        &      (fb)             &  600 (3000) fb$^{-1}$    \\      
\hline
\hline
                     1         &   0.396           &  9.002              &   3.2 (7.2)               \\ 
                     2         &   0.38           &  9.86               &   2.9 (6.6)               \\
                     3         &   0.288           &  10.915             &   2.1 (4.8)              \\ 
&&&\\  
                     $a$       &   0.39            &  8.906              &   3.2 (7.1)               \\ 
                     $b$       &   0.368           &  9.948              &   2.8 (6.4)              \\
                     $c$       &   0.295           &  10.898             &   2.2 (4.9)              \\ 
 
\hline
\hline
\end{tabular}
\caption{
Signal and total background cross sections after selection cuts for the $4b1\ell$ process 
for the benchmark points of Table~\ref{bench}, 
where the last column gives the significance 
for 600 (3000) fb$^{-1}$ integrated luminosity.
}
\label{signi}
\end{table}

The $t\bar t +$jets background cross section is normalized to the
NNLO ones by a factor $1.84$~\cite{twiki}. 
The LO $Wt$ component of the single-top cross section
is normalized to NLO by a factor 1.35~\cite{Kidonakis:2010ux}, 
while $t$- and $s$-channels by factors 1.2 and 1.47, respectively~\cite{twikisingtop}. 
The $4t$, $t\bar t h$, $t\bar t W$, $t\bar t Z$ 
cross sections at LO are adjusted to the NLO ones by factors
2.04~\cite{Alwall:2014hca}, 1.27~\cite{twikittbarh}, 1.35~\cite{Campbell:2012dh}, 1.56~\cite{Campbell:2013yla}.
The DY+jets background is normalized to  NNLO cross sections by factor 1.27~\cite{Li:2012wna,Hou:2017ozb}.
The $tWh$ and $W+$jets background are kept at LO.
The correction factors for conjugate processes are assumed to be the same for simplicity. 
Note that we do not include correction factor for the LO signal cross sections.

To distinguish signal from background, we apply 
the event selection criteria as follows.
Each event should contain one lepton, 
at least five jets, out of which at least four are $b$-tagged (denoted as $4b1\ell$).
This reduction in the required number of $b$-jets, from 
five (one from top and four from the $h$ decays) to four~\cite{Buschmann:2016uzg}, 
is in consideration of the finite $b$-tagging efficiency. 
The transverse momentum ($p_T$) of the lepton should be $> 28$ GeV, 
while $p_T > 20$ GeV for all five jets. 
The pseudo-rapidity ($\eta$) of lepton and all jets should be $|\eta|<2.5$.
We reconstruct jets by anti-$k_T$ algorithm with radius parameter $R=0.6$.
The minimum separation ($\Delta R$) between any pair of jets, 
or between the lepton and any jet, should be $> 0.4$. 
The $E_T^{\rm miss}$ is required to be $> 35$ GeV.

\begin{table}[t!]
\centering
\begin{tabular}{c |c| c| c| c | c| c| c}
\hline
&&&&&&&\\ 
      BP         & $t\bar t$   & Single     &  $t\bar t h$ & $4t$   & $t\bar t W$ &  $t\bar t Z$ & Others             \\
 &&top&&&&& \\
                  &   (fb)      &(fb)     &   (fb)       &   (fb) &   (fb)  & (fb)   & (fb)  \\      
                                
\hline
\hline
  1 & 6.701 & 1.014 & 1.008 & 0.016 & 0.022 & 0.234 & 0.007 \\  
  2 & 7.418 & 1.014 & 1.117 & 0.019 & 0.022 & 0.262 & 0.008 \\  
  3 & 7.939 & 1.521 & 1.135 & 0.024 & 0.02  & 0.268 & 0.008 \\  
&&&&&&&\\
 $a$ & 6.616 & 1.014 & 1.0   & 0.016 & 0.022 & 0.231 & 0.007 \\  
 $b$ & 7.425 & 1.014 & 1.118 & 0.019 & 0.022 & 0.262 & 0.008 \\  
 $c$ & 7.923 & 1.52  & 1.135 & 0.024 & 0.02  & 0.268 & 0.008 \\  
 
\hline
\hline
\end{tabular}
\caption{Cross sections for different background contributions after selection cuts at $\sqrt{s}=14$ TeV.}
\label{bkgcomp}
\end{table}

In order to reduce backgrounds further, we construct 
all possible $m_{bb}$ combinations from the four leading $b$-jets,
 and demand the two $m_{bb}$ pairs that are closest to $m_{h}$ 
should lie within 100 {GeV} $\leq m_{bb}\leq 150$ GeV. 
Finally, we demand the invariant mass of the four leading $b$-jets ($m_{4b}$)
to be within $|m_H-m_{4b}|< 100$ GeV. 
Note that in our exploratory study, we have not optimized the $m_{4b}$ cut 
for each of the benchmark points out of simplicity. 
We adopt the $p_T$ and $\eta$ dependent $b$-tagging efficiency 
and $c$- and light-jet misidentification efficiencies of Delphes. 
The signal and total background cross sections after selection cuts 
are summarized in Table~\ref{signi}, while 
individual components of backgrounds are given in 
in Table~\ref{bkgcomp}.

We estimate the statistical significance given in Table~\ref{signi} by use of
$\mathcal{Z} = \sqrt{2[ (S+B)\ln( 1+S/B )-S ]}$~\cite{Cowan:2010js}, where 
$S$ and $B$ are the number of signal and background events after selection cuts. 
We find that, with 600 fb$^{-1}$ data, the significance can reach above 
$\sim3.2\sigma$ for BP1 and BP$a$, $\sim 2.8\sigma$ for BP2 and BP$b$, but only $\sim 2\sigma$ for BP3 and BP$c$.
With 3000 fb$^{-1}$ at the HL-LHC, the significance  
can reach beyond $7\sigma$ for BP1 and BP$a$, about $6.5\sigma$ for BP2 and BP$b$,
and just below 5$\sigma$ for BP$3$ and BP$c$.
The significance depend heavily on the choice of $\lambda_{Hhh}$ and $\rho_{tc}$.
To get a feeling, we rescaled the significance of the BPs by $\lambda_{Hhh}= 70$ GeV 
(denoted by red dashed line in Fig.~\ref{lH0hh}) with $c_\gamma$ and $\rho_{tc}$ fixed as in 
Table~\ref{bench} and Table~\ref{branch}, respectively.
We find $\sim4.2\sigma$ is possible for BP1 and BP$a$, while $\sim 3.2\sigma$ for 
BP2 and BP$b$. The significance is below $2\sigma$ for both BP3 and BP$c$.
Note that $\lambda_{Hhh} = 70$ GeV is possible even for lower values of $c_\gamma$.
A lower $c_\gamma$ allows larger $\rho_{tc}$ for BP$a$, BP$b$ and BP$c$. 
Take $c_\gamma = 0.15$, for example, $\rho_{tc} =0.61$ is allowed, 
where one can achieve $\sim 5\sigma$, $\sim 3.8\sigma$ and $\sim 2.3\sigma$ respectively for 
BP$a$, BP$b$ and BP$c$ with 3000 fb$^{-1}$.
Though this is not as good as those shown in Table~\ref{signi},
it illustrates the chance for finding some signal for lower $\lambda_{Hhh}$ values,
but compensated by gains in $\rho_{tc}$. 
BP1, BP2, BP3 do not have this feature as discussed earlier.
In general, discovery is possible for 270 {GeV}$\lesssim m_H\lesssim330$ GeV 
with $\lambda_{Hhh} = 70$ GeV for $|m_A-m_H| < 5$ GeV, 
while significance drops for $m_H< m_{H^\pm}, m_A$ scenario.
As the parton luminosities falter away, the significance drops rapidly
if $m_H\gtrsim 340$ GeV for both scenarios.

Before closing, let us mention briefly the prospect for 
$pp\to t H \to t hh$ where $t\to b \ell^+\nu_\ell$, 
but one $h$ decays to $W^+ W^{-*}$ and the other to $b \bar b$ (conjugate process implied). 
Assuming the $W$ and $W^*$ decay leptonically, 
one has $3b3\ell$ plus $E_T^{\rm miss}$ (denoted as $3b3\ell$) signature. 
We find that discovery cannot be attained for any of the six benchmarks 
at the HL-LHC, but {$\sim 3.1\sigma$ and $\sim3\sigma$} are possible for BP1 and BP2, 
reaching{$\sim 3\sigma$ and $\sim 2.9\sigma$} significance 
for BP$a$ and BP$b$, respectively. 
The significance for BP3 is {$2.5\sigma$}, while {$2.4\sigma$} for BP$c$.
Here we follow the same cut-based analysis as described in 
Ref.~\cite{Kohda:2017fkn} for the $3b3\ell$ process, with the additional
requirement of 100 {GeV} $<m_{bb}< 150$ GeV. 
Sensitivity is poor above $m_H \gtrsim 320$ GeV,  
but if one has non-zero $\rho_{tt}$, the sensitivity to
$cg\to t hh$ is lost for $m_H > 2m_t$.
This, however, opens up the $cg\to tH\to t t \bar t$ triple-top process,
which also has $3b3\ell$ signature but
without the 100 {GeV} $<m_{bb}< 150$ GeV cut, 
which HL-LHC can actually cover~\cite{Kohda:2017fkn}.
Indeed, non-zero $\rho_{tt}$ motivates the conventional $gg\to H\to t \bar t$
search or  $gg\to H t \bar t\to t \bar t t \bar t$~\cite{Craig:2016ygr} i.e. the four-top search.
The former process suffers from large interference~\cite{Carena:2016npr} with the overwhelming $gg\to t\bar t$ background, 
however a recent search by ATLAS found some sensitivity~\cite{Aaboud:2017hnm}.
It should be clear, however, that $pp\to t H +X \to t hh$ in $3b3\ell$ 
can provide a supporting role in the top-assisted di-Higgs program 
at the HL-LHC.

\section{Discussion and summary}\label{dissc}

The 2HDM without NFC allows resonant di-Higgs production via $cg \to tH \to t h h$ process.
The process can  be searched for at the LHC via $pp\to t H +X \to t h h+X$, followed by both $h$ decays to $b\bar b$
and $t\to b\ell^+\nu_\ell$. If all other $\rho_{ij}=0$, this process 
can be discovered at HL-LHC in the mass range 270 GeV $\lesssim m_H\lesssim360$ GeV
if $\rho_{tc}\sim 0.5$ and $\lambda_{Hhh}\sim 100$ GeV. The other decay
modes such as $h h\to \gamma\gamma b \bar b$, $h h\to \tau\tau b \bar b$
are suppressed. Furthermore,  $pp\to t H +X \to t h h+X$ with $h h\to W^+W^{-*} b \bar b$ 
with $t\to b\ell^+\nu_\ell$ could be sensitive. The significances can be as large as as $\sim 2\sigma-3 \sigma$
depending on the masses of $H$, $A$ and $H^\pm$. However, both processes could be preceded by 
$pp\to t H+X \to t t \bar c$, unless $H$, $A$ are degenerate in mass and width. In such scenarios,
non-zero $\rho_{tt}$ helps via $cg\to t H\to t t \bar t$~\cite{Kohda:2017fkn}. In general,
presence of other Yukawas reduce the $H\to h h$ branching ratios, 
making discovery of top-assisted di-Higgs less likely. 
The cross section for $cg \to tH \to t h h$ vanishes as $c_\gamma$ approaches zero, and
the signature requires $c_\gamma\sim 0.15-0.2$. 
If larger $|\eta_i|$ values are allowed 
beyond $3$, $\lambda_{Hhh}$ can be enhanced even for smaller $c_\gamma$. 

Non-zero $c_\gamma$ would also induce $cg \to t H\to t W^+ W^-$ and $cg \to t H\to t ZZ$. 
We find the significances of the former process lie just below $2\sigma$ for all the benchmark points with full HL-LHC dataset. 
However, for fixed value of $c_\gamma$ and $\rho_{tc}$ a smaller $\lambda_{Hhh}$ enhances the signature for
$cg \to t H\to t W^+ W^-$ through enhanced $\mathcal{B}(H\to W^+ W^-)$.
Due to smaller $\mathcal{B}(Z\to \ell \ell)$,
we do not find $cg \to t H\to t ZZ$ to be promising for any of the benchmark points.

In summary, we have explored associated $tH \to thh$ production at the LHC
via $cg \to tH \to thh$, where production involves the extra Yukawa coupling $\rho_{tc}$,
and $H \to hh$ decay needs a finite $h$--$H$ mixing angle $\cos\gamma \neq 0$  
{as well as $\mathcal{O}(1)$ extra Higgs quartic couplings.}
We find non-negligible discovery potential at HL-LHC for $m_H \sim 300$ GeV.
Considering that $hh$ production within SM is not quite hopeful at the HL-LHC,
this is an interesting result. 
{Furthermore, a discovery might shed light on strongly first order electroweak phase transition.}
If evidence is found, not only one would have discovered New Physics induced di-Higgs production, 
but together with the companion same-sign top signal from $cg \to tH \to tt\bar c$,
one would be probing the $\rho_{tc}$ driven electroweak baryogenesis scenario
provided by this two Higgs doublet model, as well as starting to probe 
the associated Higgs potential.

\vskip0.2cm
\begin{acknowledgments}
\noindent{\bf Acknowledgments} \
We thank K.-F. Chen and Y.~Chao for fruitful discussions.
This research is supported by grants MOST 106-2112-M-002-015-MY3,
107-2811-M-002-039, and 107-2811-M-002-3069.

\end{acknowledgments}


\end{document}